\documentclass[twocolumn,prb,showpacs,preprintnumbers,amsmath,amssymb]{revtex4}

\usepackage{subfigure}
\usepackage{graphicx}
\usepackage{dcolumn}
\usepackage{bm}
\usepackage{setspace}

\begin{document}


\title{Effect of  pseudogap formation on the penetration depth of
underdoped high $T_c$ cuprates
} 

\author{J. P. Carbotte$^{1,2}$}
\author{K. A. G. Fisher$^3$}
\author{J. P. F. LeBlanc$^{3,4}$}
\author{E. J. Nicol$^{3,4}$}%
\email{nicol@physics.uoguelph.ca}
\affiliation{$^1$Department of Physics and Astronomy, McMaster
University, Hamilton, Ontario, Canada N1G 2W1}
\affiliation{$^2$The Canadian
Institute for Advanced Research, Toronto, Ontario, Canada M5G 1Z8}
\affiliation{$^3$Department of Physics, University of Guelph,
Guelph, Ontario, Canada N1G 2W1} 
\affiliation{$^4$Guelph-Waterloo Physics Institute,
University of Guelph, Guelph, Ontario, Canada N1G 2W1}

\date{\today}

\begin{abstract}
The penetration depth is calculated over the entire doping range
of the cuprate phase diagram with emphasis on the underdoped regime.
Pseudogap formation on approaching the Mott transition, for doping
below a quantum critical point, is described within a 
model based on the resonating valence bond spin liquid
which provides an ansatz for the coherent piece of the Green's function.
Fermi surface reconstruction, which is an essential element of the model,
has a strong effect on the superfluid density at $T=0$ producing
a sharp drop in magnitude, but does not
change the slope
of the linear low temperature variation. Comparison with recent
data on Bi-based cuprates provides validation of the theory and shows that
the effects of
correlations, captured by Gutzwiller factors, are essential for a 
qualitative understanding of the data. 
We find that the
Ferrell-Glover-Tinkham sum rule still holds and we
compare our results with those for the Fermi arc and the
nodal liquid models.

\end{abstract}

\maketitle

\section{Introduction}

The initial discovery of superconductivity in the cuprates
precipitated a rush to  find higher values of the
critical temperature $T_c$. In this onslaught,
not only were new superconducting members of the cuprate family
discovered but it was quickly realized that, through oxygen doping
or doping with other elements, a particular compound could display
a range of $T_c$ values. With hole-doping, in particular,
these $T_c$ values could be quite high and showed a 
dome for $T_c$ as a function  of doping in the phase diagram. While
studying optimal doping for the maximum $T_c$ became the primary focus in
initial
research, it was later appreciated that the unusual phase diagram of
these materials was of great interest in itself. Understanding
why a maximum $T_c$ exists and what controls the reduction of $T_c$
away from maximum is hoped to elucidate the physics
of the interactions
involved.  Ideally this would give direction for
what might be attempted in materials development
in order to enhance $T_c$.

Further research into the general phase diagram occurred on many
fronts, with evidence of an antiferromagnetic (AFM) insulating
state
in the parent compound and at low doping, and strange metallic
behavior
in the normal state above the superconducting dome. The most
intriguing
discovery is possibly the presence of what is now termed a
``pseudogap''
feature, occurring in the normal state on the underdoped side
of the superconducting dome, which exists for higher temperatures
as the AFM state is approached.\cite{timusk}
 This pseudogap is an energy gap-like
feature seen in normal state properties which has led to a number
of imaginative theoretical proposals for its existence. Superconductivity
in the cuprates is now thought  to have $d$-wave 
superconducting pairing, most likely due to 
spin-fluctuations.\cite{carbottenature,inversion}
It is thought, therefore, that the approach towards the AFM state
should enhance the spin-fluctuation pairing interaction $V$
 and hence increase $T_c$. However,
if this pseudogap represents a competing phase, possibly associated
with the AFM-Mott insulator, it could be responsible for a reduction
in the electronic density of states at the Fermi level
$N(0)$ which could then
reduce the superconducting $T_c$ which depends on $N(0)V$ in simple
BCS
theory. Consequently, it is natural that many
proposals for the pseudogap state have been based on a competing phase,
such as, the $d$-density wave theory.\cite{laughlin}
Others have been suggested which include the ideas of preformed 
pairs\cite{emery,randeria,alvarez,levin} 
arising
above $T_c$. 

The pseudogap phase and how it may affect
superconductivity has become a major focus of both theoretical and
experimental
work. Even knowledge of where the pseudogap line in the phase diagram
might end [possibly at $T=0$ and possibly as a quantum critical point
(QCP)]
is still a matter of debate.
Some experiments suggest it ends on the edge of the superconducting dome on the
overdoped side and others place it ending at $T=0$ inside the
superconducting
zone anywhere from optimal doping to overdoped.\cite{hufner}
 It should be noted that other works on QCPs
in
heavy fermion superconductors usually suggest such a point existing
under the superconducting dome.\cite{HF}

In spite of mounting experimental work on the underdoped cuprate
superconductors, it has been hard to develop a microscopic
theory that can build in the physics associated with the approach
from the metallic state to the Mott insulator and give a workable
formalism, which includes doping,
 for providing theoretical insights and predictions for
experiments both in the normal and superconducting state (including
the idea of a pseudogap). Further to this issue has been the more
recent experimental development where a possible reconstruction of 
the Fermi surface in the underdoped cuprates
has been found to occur based on seeing Fermi 
pockets from de Haas-van Alphen experiments\cite{louis} and
or Fermi arcs in  angle-resolved photoemission
(ARPES) experiments.\cite{arpesarcs} There has also been one report
of an observation of pockets from ARPES.\cite{zhou}
Moreover, there has
been indirect evidence of arcs from other
experiments, such as optical spectroscopy\cite{hwang}, specific
heat\cite{storey} and scanning tunneling spectroscopy\cite{davis}
 measurements. Indeed,
attempting to resolve the arc versus pocket debate has been the
subject of numerous papers.

One theory which is promising in this regard has been due to
Yang, Rice and Zhang\cite{yrz} (YRZ) which is based on previous studies of
a resonating valence bond (RVB) spin liquid state originally proposed
by Anderson\cite{anderson}. 
The merit of the YRZ approach is that they have used
previous numerical and theoretical RVB-type studies to develop an
ansatz for the many-body Green's function that would represent
the RVB spin liquid. This ansatz builds in the approach to the
AFM-Mott 
insulator and shows a reconstructed Fermi surface which is a
large
Fermi surface when there is no pseudogap, above optimal doping,
but is reconstructed to small Fermi or Luttinger pockets (which look more like
arcs when the quasiparticle weight is included) for the underdoped
case.
The pseudogap opens up around the AFM Brillouin zone in this theory
and it gaps out or reconstructs part of the Fermi surface. So
far this theory has been used to evaluate a number of experimental
properties\cite{belen,emilia,yrzarpes,james} 
with good qualitative agreement and, in the same spirit as
BCS theory, using this ansatz allows us to determine what are the
essential
elements that should go into a more sophisticated microscopic theory
should it be developed in the future.

In this work, we wish to examine the long-standing puzzle associated
with
the penetration depth measurements. The penetration depth
was one of the first experiments to clearly indicate that a $d$-wave
order parameter symmetry for the superconductivity was present in
the cuprates.\cite{bonnhardy} This was an extremely important result
in influencing the direction of research in this field as it allowed
for the elimination of a number of possible mechanisms for Cooper pairing.
It also clarified the need for very high quality samples to remove
the obscuring features due to impurities.
The essential observation from the experiment was that the
superfluid density $\rho_s(T)$, which  is related to the penetration
depth $\lambda(T)$ by $\rho_s(T)\propto 1/\lambda^2(T)$, showed
a low temperature linear 
$T$ behavior as expected for a clean BCS $d$-wave superconductor,
i.e., $\rho_s(T)=\rho_s(0)-bT$. However,
in the underdoped regime, it is known\cite{uemura,leeandwen,franz} 
that while the zero temperature
value of the superfluid density depends strongly on the doping $x$,
the coefficient $b$ of the first linear-in-$T$ correction is much less sensitive
to $x$. This result cannot be understood within a simple BCS $d$-wave model.
Our goal in this paper is to study the penetration depth in the YRZ
model
and to see if the experimental data, with its doping dependence, can
be explained by this theory. Furthermore, we wish to see if there is
evidence for a reconstructed Fermi surface in the penetration depth
data.
With this study we can develop a better understanding of the physics
which is giving rise to this non-BCS behavior of the doping dependence
of the penetration depth.

Our paper is structured as follows.
In Section~II, we introduce the basic features of the YRZ theory
that enter into our calculations. This is followed by a discussion
of the penetration depth formula used in this work and its various
limits, given in Section~III. In Section~IV, we summarize some
theoretical
formulas associated with the frequency-dependent optical conductivity, 
a quantity
which is also related to the penetration depth. This will aid in our
discussion of the Drude weight and the sum rule. We then present our
results in Section~V and provide our conclusions in Section~VI.

\section{Theoretical model of YRZ}

\begin{figure}
\includegraphics[width=0.45\textwidth]{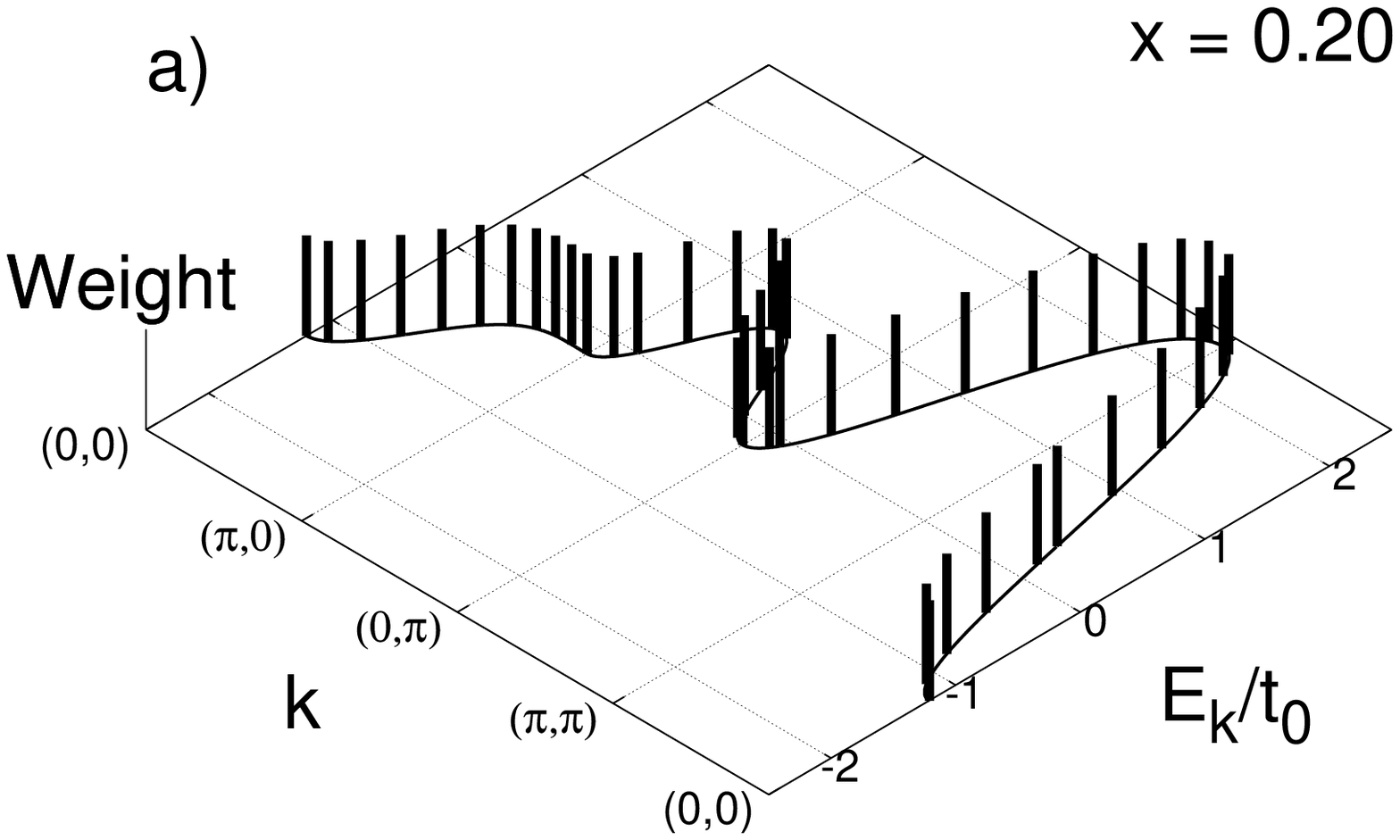}
\includegraphics[width=0.45\textwidth]{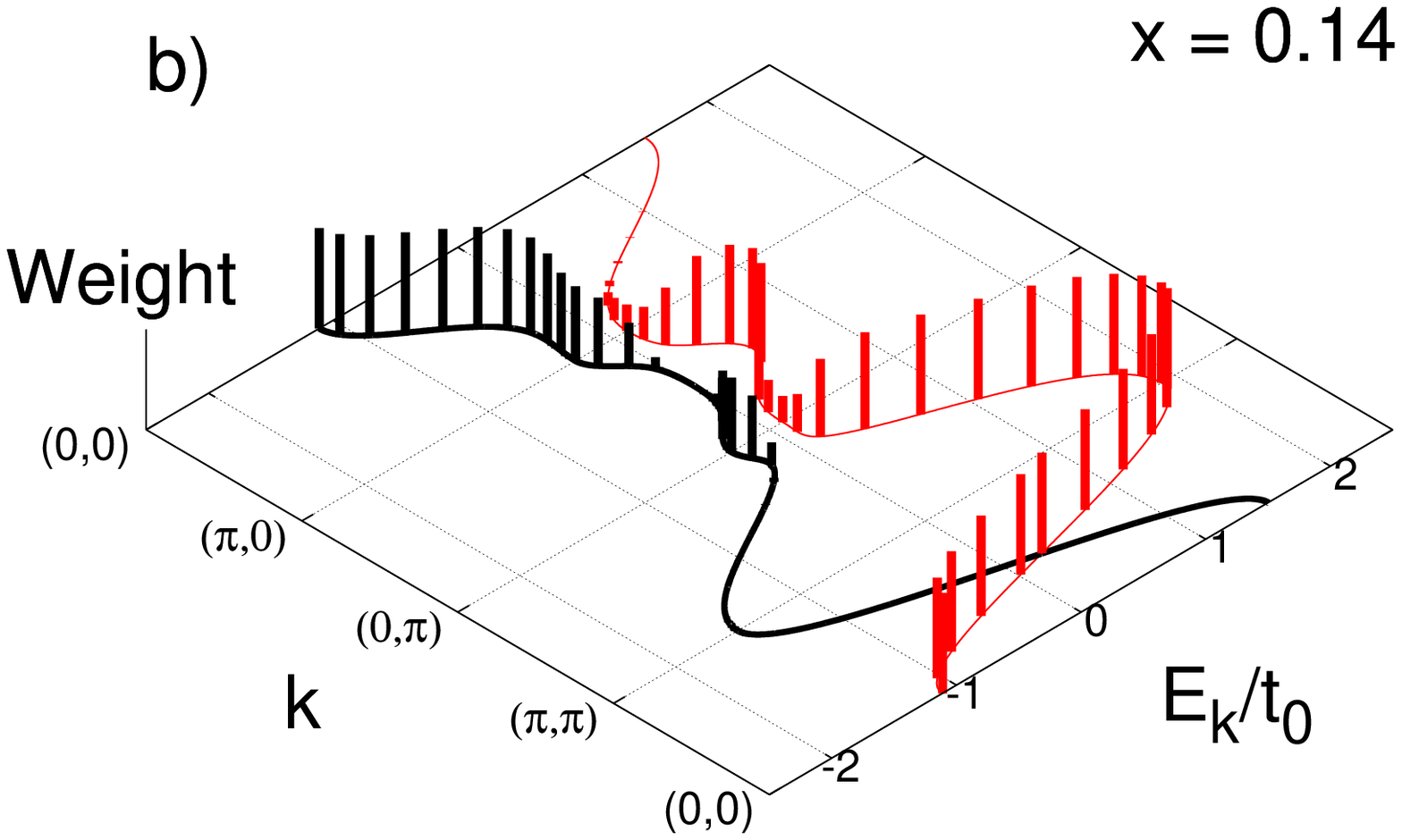}
\caption{\label{fig1} (Color online) The electronic band structure
for (a) $x=0.2$ with no pseudogap and for (b) $x=0.14$ with a
pseudogap.
In the YRZ model for the pseudogapped normal state, two bands,
$E_{\boldsymbol{k}}^+$ and $E_{\boldsymbol{k}}^-$, exist
with ${\boldsymbol{k}}$-dependent weighting factors $W_{\boldsymbol{k}}^+$ and
$W_{\boldsymbol{k}}^-$, respectively, shown as a peak of height set by the weight.
The $E_{\boldsymbol{k}}^-$ band is the lower energy band shown in heavy black and
the $E_{\boldsymbol{k}}^+$ band is the higher energy band shown in lighter red.
}
\end{figure}

The YRZ model provides an ansatz for the coherent part of
the  many-body Green's function for the case of a doped RVB spin liquid.
It includes a dependence on doping $x$ and is given as:\cite{yrz,yrzarpes}
\begin{equation}
G(\boldsymbol{k},\omega,x)=
\sum_{\alpha=\pm}  \frac{g_tW^{\alpha}_{\boldsymbol{k}}}
{\omega-E^{\alpha}_{\boldsymbol{k}}-\Delta_{\rm sc}^{2}/(\omega+E^{\alpha}_{\boldsymbol{k}})},
\label{eq:G}
\end{equation}
where 
\begin{eqnarray}
E_{\boldsymbol{k}}^ \pm  &=& \frac{{\xi_{\boldsymbol{k}}  - \xi_{\boldsymbol{k}}^0 }}{2} \pm E_{\boldsymbol{k}},\nonumber\\
E_{\boldsymbol{k}} &=& \sqrt {\tilde{\xi_{\boldsymbol{k}}}^2  + \Delta
  _{\rm pg}^2 },\nonumber\\
\tilde{\xi_{\boldsymbol{k}}} & =& \frac{(\xi_{\boldsymbol{k}}  + \xi_{\boldsymbol{k}}^0 )}{2},\nonumber\\
W_{\boldsymbol{k}}^ \pm   &=& \frac{1}{2}\left( {1 \pm
  \frac{{\tilde{\xi_{\boldsymbol{k}}}  }}{E_{\boldsymbol{k}}}}
\right).
\end{eqnarray}
 In these expressions 
$\xi_{\boldsymbol{k}}^0  =  - 2t(x)(\cos k_xa  +
\cos k_ya)$ is the first nearest neighbor tight-binding dispersion, 
which introduces the 
antiferromagnetic Brillouin zone boundary (AFBZ) into the Green's
function along which the pseudogap $\Delta_{\rm pg}$ develops. The band structure
$\xi_{\boldsymbol{k}} = - 2t(x)(\cos
k_xa  + \cos k_ya ) - 4t^{\prime}(x) \cos k_xa \cos k_ya - 2t''(x)(\cos
2k_xa  + \cos 2k_ya )-\mu_p$  is taken from tight-binding for a 
 system which includes hopping terms up to third
nearest neighbor, with $\mu_p$, a chemical potential determined
by the Luttinger sum rule.\cite{yrz}
Doping dependence enters in three ways
into the Green's function. First, the band structure is doping dependent
through the 
hopping coefficients: $t(x)=g_{t}(x)t_{0}+3g_{s}(x)J\chi/8$, $t^{\prime
}(x)=g_{t}(x)t_{0}^{\prime }$, and $t^{\prime\prime
}(x)=g_{t}(x)t_{0}^{\prime \prime }$,  where $g_{t}(x)=2x/\left(
1+x\right)$ and $g_{s}(x)=4/(1+x)^{2}$ are the Gutzwiller factors
and $J/t_{0}=1/3$ and $\chi=0.338$. 
YRZ use $t^{\prime}_0/t_{0}=-0.3$, $t^{\prime\prime}_0/t_{0}=0.2$, 
a choice of parameters 
to match this energy dispersion to that calculated for
Ca$_2$CuO$_2$Cl$_2$ (Ref.~\cite{yrz}). 
The $x$-dependence of these coefficients reflects the fact that
strong correlations will narrow the bands as the Mott insulator is
approached. Second, the coherent part of the Green's function 
changes with doping with a
Gutzwiller factor given by $g_t\equiv g_t(x)$. Such factors reflect
the projection out of doubly occupied states and the approach to the
atomic limit.
Third, the magnitude of the 
pseudogap, $\Delta_{\rm pg}$, and superconducting gap, $\Delta_{\rm sc}$,
are also doping dependent, as inferred from experiment and the behavior of
$T_c$. These gaps are 
taken to be $d$-wave, such that we can write them as 
$\Delta_{\rm pg}=\frac{\Delta_{\rm pg}^{0}(x)}{2}(\cos
k_xa -\cos k_ya)$ and  $\Delta_{\rm sc}=\frac{\Delta_{\rm
    sc}^{0}(x)}{2}(\cos k_xa -\cos k_ya)$, respectively,
 with $a$ the lattice constant, $\Delta^{0}_{\rm
    pg}(x)/t_0=0.6(1-x/0.2)$
and $\Delta^{0}_{\rm
    sc}(x)/t_0=0.14(1-82.6(x-0.2)^2)$. In general, the gap could
contain many higher harmonics\cite{branch1995} as is also the case
in conventional superconductors\cite{tomlinson,leung1,leung2} but such
complications are not essential for a first understanding.
From Eq.~(\ref{eq:G}), one can extract the YRZ 
 spectral function $A(\boldsymbol{k},\omega)$
 and see that there are 
four energy branches, given by the energies, $\pm E_{S}^{\alpha}$,
where  $E^{\alpha}_S=\sqrt{(E_{\boldsymbol{k}}^{ \alpha })^2  +
  \Delta _{\rm sc}^2 }$. 
Following the usual development of the 
equations of 
superconductivity, both the  regular and anomalous spectral functions
are found to be:
\begin{eqnarray}
A(\boldsymbol{k},\omega)&=&\sum_{\alpha=\pm}g_tW_{\boldsymbol{k}}^\alpha
[(u^\alpha)^2\delta(\omega-E^\alpha_S)+(v^\alpha)^2\delta(\omega+E^\alpha_S)],
\nonumber\\
\\
B(\boldsymbol{k},\omega)&=&\sum_{\alpha=\pm}g_tW_{\boldsymbol{k}}^\alpha
u^\alpha v^\alpha[\delta(\omega-E^\alpha_S)-\delta(\omega+E^\alpha_S)],
\end{eqnarray}
where 
\begin{eqnarray}
u^\alpha&=&\biggl[\frac{1}{2}\biggl(1+\frac{E_{\boldsymbol{k}}^\alpha}{E^\alpha_S}\biggr)\biggr]^{1/2},\\
v^\alpha&=&\biggl[\frac{1}{2}\biggl(1-\frac{E_{\boldsymbol{k}}^\alpha}{E^\alpha_S}\biggr)\biggr]^{1/2}.
\end{eqnarray}
These two spectral functions enter the expressions for the penetration
depth and the optical conductivity which are described in the
following
sections. In the limit of $\Delta_{\rm sc}=0$ but with a finite pseudogap,
there are still
two branches in the quasiparticle energy dispersion as shown in Fig.~\ref{fig1}(b)
for various symmetry directions in the first quadrant of the square
Brillouin zone. 
The weighting factors $W_{\boldsymbol{k}}^\alpha$ are shown as upright bars
of varying height in the third dimension. Comparing with a case for no
pseudogap where there  is just the one energy dispersion with full weight
[Fig.~\ref{fig1}(a)], we see that
a gap in energy due to the pseudogap formation has opened around 
the antinodal direction in the Brillouin zone quadrant at $(\pi,0)$ and 
$(0,\pi)$, and
along the AFBZ from $(\pi,0)$ to $(0,\pi)$. The presence of the two
bands gives rise to interband transitions in the frequency-dependent
optical conductivity
and a depletion of the intraband Drude component as we will later discuss.

In the following work, we will also examine several other models
in relation to the YRZ model. We will refer to a Fermi liquid which is
just the case of taking the pseudogap to be zero in our formalism. We also
use the term YRZ modified (YRZ mod.).
In the YRZ model, the presence of the pseudogap reconstructs the large
Fermi surface (found on the overdoped side of the phase diagram) into
a small Fermi pocket. If we use the YRZ model but take the 
superconducting $d$-wave gap to be finite 
only in the region of the Fermi pocket and to be zero beyond the pocket 
towards the antinodal direction, then we will call this YRZ 
modified.
In conventional superconductors\cite{tomlinson,leung1,leung2}
such as Pb and Al, there are directions where the
Fermi surface is gapped out by the crystal potential and one finds that
the superconducting gap is also zero there. In the present case it is
the pseudogap which prevents the superconducting gap from having its
full amplitude 
in certain regions.

Two other models in the literature are the nodal liquid and the Fermi arc
model. In these two models, the pseudogap is taken to be on the large
Fermi 
surface given by $\xi_{\boldsymbol{k}}$. In the nodal liquid,
 both the pseudogap
and superconducting gaps are active over the entire Fermi surface.
In the arc model, the pseudogap  is only finite in a
region of the Fermi surface near that antinodal direction on an
arc of the large Fermi surface that is defined by a 
critical angle $\theta_c$
measured at the $(\pi,\pi)$ point from the Brillouin zone boundary towards the
nodal direction.\cite{james}

\section{Penetration Depth}

The London penetration depth $\lambda(T)$ is given by the zero
frequency limit of the imaginary part of the optical conductivity
$\sigma(T,\Omega)$ at temperature $T$. Specifically,
\begin{equation}
\frac{1}{\lambda^2(T)}=\displaystyle\lim_{\Omega\to 0}\frac{4\pi}{c^2}\Omega\rm{Im}
\sigma(T,\Omega),
\end{equation}
where $c$ is the velocity of light. The imaginary part of $\sigma(T,\Omega)$
has both a contribution from the paramagnetic and diamagnetic part of the
current. A particularly transparent formal expression, which manifestly
vanishes when the superconducting gap vanishes, results when both contributions
are treated on the same footing\cite{branch,radtke}
\begin{eqnarray}
&&\frac{1}{\lambda^2(T)}=\displaystyle\lim_{\Omega\to 0}\frac{16\pi e^2}{c^2}
\sum_{\boldsymbol{k}} v_{k_x}^2\nonumber\\
&\times&\int d\omega' d\omega'' \lim_{\boldsymbol{q}\to0}
\biggl(\frac{f(\omega'')-f(\omega')}{\omega''-\omega'}\biggr)
B(\boldsymbol{k}+\boldsymbol{q},\omega')B(\boldsymbol{k},\omega''),\nonumber\\
\label{eq:BB}
\end{eqnarray}
where $e$ is the electron charge and $f(\omega)$ is the Fermi function 
$1/[1+\exp(\beta\omega)]$, with $\beta=1/(k_BT)$ and $k_B$ the
Boltzmann constant.
$B({\boldsymbol{k}},\omega)$ is the anomalous spectral density previously introduced and is
proportional to the superconducting gap. Consequently, this expression vanishes in the
normal state as it must. After some lengthy but straightforward algebra,
one arrives at a more explicit formula for $1/\lambda^2(T)$ in the YRZ model
of the form 
\begin{eqnarray}
\frac{1}{\lambda^2(T)}&=&\frac{16\pi e^2}{c^2}\sum_{\boldsymbol{k}}
 v_{k_x}^2\nonumber\\
&\times &\biggl[(W_{\boldsymbol{k}}^{+})^2(u^+v^+)^2\biggl(2\frac{\partial f(E^+_S)}{\partial E^+_S}+\frac{1-2f(E_S^+)}{E_S^+}\biggr)\nonumber\\
&+&(W_{\boldsymbol{k}}^{-})^2(u^-v^-)^2\biggl(2\frac{\partial f(E^-_S)}{\partial E^-_S}+\frac{1-2f(E_S^-)}{E_S^-}\biggr)\nonumber\\
&+&4W_{\boldsymbol{k}}^+W_{\boldsymbol{k}}^-u^+v^+u^-v^-\biggl(\frac{f(E_S^+)-f(E_S^-)}{E_S^+-E_S^-}\nonumber\\
&&\qquad
+\frac{1-f(E_S^+)-f(E_S^-)}{E_S^++E_S^-}\biggr)\biggr].
\label{eq:yrzpen}
\end{eqnarray}
This formula is for the clean limit and ignores the inelastic effects
which are not large for the penetration depth.\cite{ewald} Only the
coherent
part of the Green's function contributes to the condensate. Note that
we have suppressed for now the factor of $g_t^2$ which weights this
quantity in the YRZ model. We will be presenting all of our numerical
work for $1/\lambda^2(T)$ in units of $e^2t_0/\hbar^2 d$, 
where $d$ is the $c$-axis
distance
per Cu-O plane (not shown explicitly in the superfluid density
formulas given here which are written for 2 dimensions).

Our basic formula, Eq.~(\ref{eq:yrzpen}), can be reduced  in simpler
models such as the nodal liquid or the usual Fermi arc model. In both cases,
the pseudogap is placed on the Fermi surface. For the arc model it is
nonzero only on arcs about the antinodal direction. To place the
pseudogap on the Fermi surface instead of having it on the antiferromagnetic
Brillouin zone, we take $\xi_{\boldsymbol{k}}^0$ to be 
equal to $\xi_{\boldsymbol{k}}$ so that $E_{\boldsymbol{k}}^\pm=\pm E_{\boldsymbol{k}}$,
 and $W_{\boldsymbol{k}}^{\pm}=(1\pm\xi_{\boldsymbol{k}}/E_{\boldsymbol{k}})/2$, with $E_{\boldsymbol{k}}=\sqrt{\xi_{\boldsymbol{k}}^2+\Delta_{\rm pg}^2}$.
It follows that $E_S^\pm=\sqrt{\xi_{\boldsymbol{k}}^2+\Delta_{\rm pg}^2+\Delta_{\rm sc}^2}\equiv E_S$,
$W_{\boldsymbol{k}}^++W_{\boldsymbol{k}}^-=1$,  $(u^\pm)^2=(1+E_{\boldsymbol{k}}/E_S)/2\equiv u^2$,
$(v^\pm)^2=(1-E_{\boldsymbol{k}}/E_S)/2\equiv v^2$, and 
$u^\pm v^\pm=uv=\Delta_{\rm sc}/2E_S$. We then obtain
\begin{eqnarray}
\frac{1}{\lambda^2(T)}&=&\frac{16\pi e^2}{c^2}\sum_{\boldsymbol{k}} v_{k_x}^2 (uv)^2\biggl(
2\frac{\partial f(E_S)}{\partial E_S}+\frac{1-2f(E_S)}{E_S}\biggr)\nonumber\\
&=&\frac{4\pi e^2}{c^2}\sum_{\boldsymbol{k}} v_{k_x}^2 \frac{\Delta_{\rm sc}^2}{E_S^2}\biggl\{
-\frac{\partial}{\partial E_S}+\frac{1}{E_S}\biggr\}{\rm tanh}\biggl(\frac{\beta
E_S}{2}\biggr).\nonumber\\
\label{eq:bpen}
\end{eqnarray}
If we further take the pseudogap to be zero $E_S=\sqrt{\xi_{\boldsymbol{k}}^2+\Delta_{\rm sc}^2}$
then Eq.~(\ref{eq:bpen}) becomes a well-known formula for the Fermi
liquid
(FL) penetration
depth in BCS theory.\cite{donovan1,donovan2,donovan3,franz,misawa} With the 
pseudogap non-zero, we will refer to Eq.~(\ref{eq:bpen}) as the nodal liquid limit
and if the pseudogap is cut off at a certain critical angle $\theta_c$
 away from the antinodal
direction so that there is no pseudogap on an arc about the nodal direction,
we will refer to this as the arc model.

Examining various limits of Eq.~(\ref{eq:bpen}) provides understanding
of the physics when a pseudogap is present. We recall that for the BCS
case with no pseudogap, at zero temperature the tanh$(\beta E/2)$ is equal to 1
and the $T=0$ limit of the penetration depth takes on a particularly
simple form
\begin{eqnarray}
\frac{1}{\lambda^2(0)}&=&\frac{2\pi e^2}{c^2}N(0)v_F^2\nonumber\\
&\times&\int_0^{2\pi}\frac{d\theta}{2\pi}\int_{-\infty}^{+\infty} d\epsilon
\frac{[\Delta^{0}_{\rm sc}\cos(2\theta)]^2}{\{\epsilon^2+[\Delta^{0}_{\rm sc}\cos(2\theta)]^2\}^{3/2}},\nonumber\\
\label{eq:pencont}
\end{eqnarray}
where the continuum limit of the free electron bands has been taken to simplify
the mathematics and $N(0)$ is the electronic density of states, assumed
to be constant. The ratio of the electron density to the electron mass 
$n/m=N(0)v_F^2$. Further, the last integral over energy in Eq.~(\ref{eq:pencont})
is independent of the factor of $[\Delta_{\rm sc}^0\cos(2\theta)]$ and equal
to 2. The angular integration normalized to $2\pi$ is then trivially equal
to 1. Thus we obtain the well-known result that $1/\lambda^2(0)=4\pi n e^2/mc^2$.
Moreover, in the limit $T\to 0$, the leading temperature dependence is given by
the derivative term $(-\partial/\partial E)$ in Eq.~(\ref{eq:bpen}). If we write
for simplicity $\Delta_S\equiv\Delta^0_{\rm sc}\cos(2\theta)$, we arrive at
\begin{equation}
\frac{\lambda^2(0)}{\lambda^2(T)}=1-\beta
\int_0^{2\pi}\frac{d\theta}{2\pi}\int_{-\infty}^{+\infty} d\epsilon
\frac{\Delta^2_S}{\epsilon^2+\Delta^2_S}\frac{e^{\beta\sqrt{\epsilon^2+\Delta^2_S}}}{[e^{\beta\sqrt{\epsilon^2+\Delta^2_S}}+1]^2}.
\label{eq:penlowT}
\end{equation}
 As $T\to 0$ this last integral is strongly 
peaked about $\epsilon\sim 0$ and $\Delta_S\sim 0$ (i.e., the nodal direction)
and we obtain the standard result
\begin{equation}
\frac{1}{\lambda^2(T)}=\frac{4\pi ne^2}{mc^2}\biggl[1-2\ln 2\frac{k_BT}{\Delta^0_{\rm sc}}\biggr].
\end{equation}
For the case of the nodal liquid, the pseudogap is assumed to go like 
$\Delta_{\rm pg}(\theta)=\Delta^0_{\rm pg}\cos(2\theta)$ over the entire Fermi
surface as does the superconducting gap. In this case, expression
(\ref{eq:bpen}) can be cast in the form of the standard BCS case
with two changes. The
square of the gap amplitude is to be replaced by the sum of the squares
of superconducting and pseudogap, i.e., $\Delta_{\rm sc}^0\to\sqrt{(\Delta_{\rm sc}^0)^2+
(\Delta_{\rm pg}^{0})^2}$,
and an overall factor of $(\Delta_{\rm sc}^{0})^2/[(\Delta_{\rm sc}^{0})^2+(\Delta_{\rm pg}^{0})^2]$
now multiplies the entire expression. This leads immediately to the result
\begin{eqnarray}
\frac{1}{\lambda^2(T)}&=&\frac{4\pi ne^2}{mc^2}
\frac{(\Delta^{0}_{\rm sc})^2}{(\Delta_{\rm sc}^{0})^2+(\Delta_{\rm pg}^{0})^2}\nonumber\\
&\times&\biggl[1-2\ln 2\frac{k_BT}{\sqrt{(\Delta_{\rm sc}^{0})^2+(\Delta_{\rm pg}^{0})^2}}\biggr].
\label{eq:pen4}
\end{eqnarray}
In this case the London penetration depth at $T=0$ is no longer simply a normal
state property but depends explicitly 
on the value of the superconducting gap amplitude.
Also, the normalized slope of the linear-in-temperature term is greatly
reduced when the pseudogap is large. 
While this limit is helpful because of its simplicity,
we will see that it does not describe the YRZ results well.
By contrast, the arc model which is closely related to the above
equations does well in capturing the main physics contained in the more
mathematically complex YRZ model. Before turning to this case, we note that
at $T=0$, when the pseudogap is small compared with the superconducting gap,
the penetration depth is modified to 
$1/\lambda^2(0)=(4\pi ne^2/mc^2)[1-(\Delta^0_{\rm pg}/\Delta^0_{\rm sc})^2]$
which tells us that the superfluid density is effectively reduced over
its no pseudogap value. The pseudogap competes with the superconductivity
for phase space. 

Introduction of an arc over which the pseudogap is
zero, while finite from antinodal direction to $\theta_c$, modifies Eq.~(\ref{eq:pencont})
but at the same time leaves the second integral in Eq.~(\ref{eq:penlowT}) 
completely unaltered because, for sufficiently small temperature, only the
angular region very close to the nodal direction is of importance. But in
these regions, the pseudogap is zero so that the first derivative
term $-\partial/\partial E$ in Eq.~(\ref{eq:bpen}) is unaffected by the
pseudogap. Therefore, we obtain for this term
$(4\pi ne^2/mc^2)[-(2\ln 2)k_BT/\Delta^0_{\rm sc}]$ which is completely
unchanged from its Fermi liquid value. This is not so for the value
of  the zero
temperature penetration depth. This quantity does know about the pseudogap.
In the continuum approximation, it is given by
\begin{eqnarray}
&&\frac{1}{\lambda^2(0)}=\frac{2\pi ne^2}{mc^2}\nonumber\\
&\times&\int_0^{2\pi}\frac{d\theta}{2\pi}\int_{-\infty}^{+\infty} d\epsilon
\frac{[\Delta^{0}_{\rm sc}\cos(2\theta)]^2}{\{\epsilon^2+
[(\Delta^{0}_{\rm sc})^2+(\bar\Delta^{0}_{\rm pg})^2]\cos^2(2\theta)\}^{3/2}},
\nonumber\\
\end{eqnarray}
where for simplicity we have assumed the pseudogap to have the same 
angular dependence as the superconducting gap, namely the $d$-wave
$\cos(2\theta)$ dependence. Here the bar on $\bar\Delta^0_{\rm pg}$ is to mean
that it is zero in the interval $\theta_c$ to $\pi/4$ and all other
symmetry related intervals. Next we note that the energy integral will give
$2(\Delta^{0}_{\rm sc})^2/[(\Delta_{\rm sc}^{0})^2+(\Delta_{\rm pg}^{0})^2]$ from the regions where 
$\bar\Delta_{\rm pg}^0$ is nonzero (antinodal) and 2 from the regions where 
$\bar\Delta_{\rm pg}^0$ is zero (nodal). We finally get 
\begin{equation}
\frac{1}{\lambda^2(0)}=\frac{4\pi ne^2}{mc^2}\biggl[1-\frac{4\theta_c}{\pi}
\frac{(\Delta^{0}_{\rm pg})^2}{(\Delta_{\rm sc}^{0})^2+(\Delta_{\rm pg}^{0})^2}\biggr].
\label{eq:pen6}
\end{equation}
For $\theta_c=0$ (that is, no pseudogap), we get back the classical
London result of Eq.~(\ref{eq:pencont}) after integration over energy and
angle. For  $\theta_c=\pi/4$, the nodal liquid result of Eq.~(\ref{eq:pen4}) 
in the 
$T\to 0$ limit is obtained. When the pseudogap $\Delta_{\rm
  pg}^0=0$,
we recover the known result of Fermi liquid theory, that the zero
temperature
value of the penetration depth depends only on normal state parameters
and not on the explicit value of the superconducting gap itself. When
the pseudogap is small as compared to the superconducting gap, the
correction to 1 in Eq.~(\ref{eq:pen6})
is small and of order $(4\theta_c/\pi)(\Delta_{\rm
  pg}^0/\Delta_{\rm sc}^0)^2$. In the
opposite
limit of large pseudogap, the superconducting gap drops out and the
correction is $4\theta_c/\pi$, explicitly independent of both gaps.
The pseudogap, however, implicitly determines the critical angle
$\theta_c$ related to the part of the Fermi surface over which
the pseudogap is non-zero. 

\section{Optical Conductivity}

The real part of the optical conductivity, ${\rm Re}\sigma(T,\Omega)$,
is given by\cite{emilia}
\begin{widetext}
\begin{equation}
{\rm Re}\sigma(T,\Omega)=\frac{2\pi e^2}{\Omega}
\sum_{\boldsymbol{k}} v_{k_x}^2\int_{-\infty}^{+\infty} 
d\omega [f(\omega) -f(\omega+\Omega)]
[A({\boldsymbol{k}},\omega)A({\boldsymbol{k}},\omega+\Omega)+B({\boldsymbol{k}},\omega)B({\boldsymbol{k}},\omega+\Omega)].
\label{eq:cond1}
\end{equation}
In the clean limit of the YRZ model, we obtain, after long but
straightforward
algebra, the expression
\begin{eqnarray}
{\rm Re}\sigma(T,\Omega)
&=&2\pi e^2
\sum_{\boldsymbol{k}} v_{k_x}^2
\biggl\{\delta(\Omega)\biggl[-\frac{\partial
  f(E^+_S)}{\partial E^+_S}(W_{\boldsymbol{k}}^+)^2-\frac{\partial f(E^-_S)}{\partial
  E^-_S}(W_{\boldsymbol{k}}^-)^2\biggr]\nonumber\\
&+&W_{\boldsymbol{k}}^+W_{\boldsymbol{k}}^-\biggl[(u^-v^+-u^+v^-)^2\frac{1-f(E_S^+)-f(E_S^-)}{E_S^++E_S^-}
\delta(\Omega-E_S^+-E_S^-)\nonumber\\
&-&(u^+u^-+v^+v^-)^2
\frac{f(E_S^+)-f(E_S^-)}{E_S^+-E_S^-}[\delta(\Omega-E_S^++E_S^-)
+\delta(\Omega+E_S^+-E_S^-)]
\biggr]\biggr\}.\nonumber\\
\label{eq:cond2}
\end{eqnarray}
\end{widetext}
The first term in Eq.~(\ref{eq:cond2}) is proportional to
$\delta(\Omega)$
and a Drude weight can be defined as
\begin{eqnarray}
\frac{W_D}{2}&\equiv&\int_{0}^\infty{\rm Re}\sigma_{\rm
    Drude}(T,\Omega)d\Omega\nonumber\\
&=&2\pi e^2
\sum_{{\boldsymbol{k}},\alpha=\pm} v_{k_x}^2\biggl[-\frac{\partial
  f(E^\alpha_S)}{\partial E^\alpha_S}(W_{\boldsymbol{k}}^\alpha)^2\biggr].
\label{eq:cond3}
\end{eqnarray}
The second piece is the interband contribution because
it
involves transitions between $E_S^+$ and $E_S^-$, and after
integration
over $\Omega$ we have its weight $W_{IB}$ equal to 
\begin{widetext}
\begin{eqnarray}
W_{IB}&\equiv&\int_{0}^\infty{\rm Re}\sigma_{\rm Interband}(T,\Omega)d\Omega\nonumber\\
&=&2\pi e^2
\sum_{\boldsymbol{k}} v_{k_x}^2
W_{\boldsymbol{k}}^+W_{\boldsymbol{k}}^-\biggl[(u^-v^+-u^+v^-)^2\frac{1-f(E_S^+)-f(E_S^-)}{E_S^++E_S^-}\nonumber\\
&&\qquad
-(u^+u^-+v^+v^-)^2
\frac{f(E_S^+)-f(E_S^-)}{E_S^+-E_S^-}
\biggr].
\label{eq:cond4}
\end{eqnarray}
\end{widetext}
These two quantities are closely related to the penetration depth.
As in that case, an interesting limit of these expressions occurs
when the pseudogap is placed on the Fermi surface, {\it i.e.}, taking
$\xi_0=\xi_{\boldsymbol{k}}$, the same algebraic simplifications as previously described
apply. First it is instructive to look at Eq.~(\ref{eq:cond2}) before
integration over photon energy $\Omega$. The second term vanishes
because  the combination of $u$'s and $v$'s in the small square
bracket vanishes and the third (last) term becomes 
$2W_{\boldsymbol{k}}^+W_{\boldsymbol{k}}^-[-\partial f(E_S)/\partial E_S]\delta(\Omega)$ and combines with the
first
term to give
\begin{equation}
{\rm Re}\sigma(T,\Omega)=2\pi e^2\sum_{\boldsymbol{k}}v^2_{k_x}
\biggl[-\frac{\partial
  f(E_S)}{\partial E_S}\biggr]\delta(\Omega),
\label{eq:cond5}
\end{equation}
where we have noted that $W_{\boldsymbol{k}}^++W_{\boldsymbol{k}}^-=1$. 
In the
on-the-Fermi-surface 
limit of the pseudogap, the sum of the two
contributions
to the total weight can still be denoted by $W_D/2$ as before and
\begin{equation}
\frac{W_D}{2}=2\pi e^2\sum_{\boldsymbol{k}}v^2_{k_x}
\biggl[-\frac{\partial
  f(E_S)}{\partial E_S}\biggr].
\label{eq:cond6}
\end{equation}
In the continuum or free electron model for the band structure,
with neither superconductivity nor pseudogap and at zero temperature,
$W_D=n\pi e^2/m\equiv\Omega_p^2/4$, a well known result where $\Omega_p$
is the normal state plasma frequency.
For superconductivity over the entire Fermi
surface, but no pseudogap, we get zero as expected. In the clean limit at $T=0$,
the entire real part of the conductivity goes into the condensate.
This would also hold for the nodal liquid.
However, in the arc model with no
superconductivity,
we have a finite $W_D$, namely $[n\pi e^2/m](1-4\theta_c/\pi)$
corresponding to the gapless arc that remains about the nodal
direction.
The limit $\theta_c=0$ gives no reduction from its Fermi liquid value
and $\theta_c=\pi/4$ corresponds to a fully gapped Fermi surface (nodal
liquid).

In conventional BCS theory, the Ferrell-Glover-Tinkham\cite{FGT1,FGT2}
sum rule
states that the optical spectral weight is not changed on entering
the superconducting state. In terms of the notation introduced here,
it
reads
\begin{equation}
\frac{c^2}{8\lambda^2(T)}+\int_{0^+}^\infty d\Omega 
{\rm  Re}\sigma_S(T,\Omega)= \int_{0}^\infty d\Omega 
{\rm  Re}\sigma_N(T,\Omega),
\label{eq:cond7}
\end{equation}
where we denote by $S/N$ the superconducting/normal state,
respectively.
\begin{figure}
\includegraphics[width=0.45\textwidth]{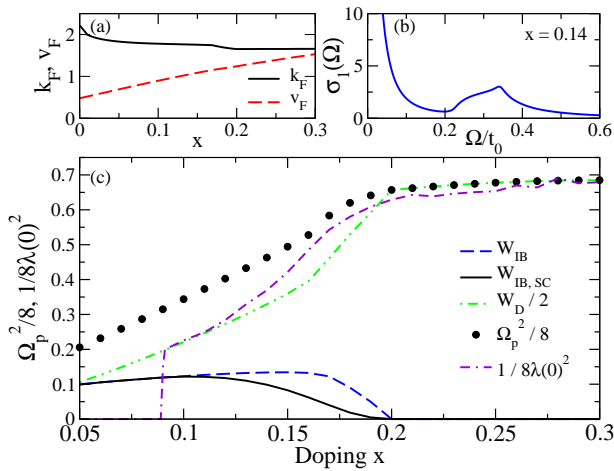}
\caption{\label{fig2} (Color online) (a) The Fermi wavevector
$k_F$ and velocity $v_F$ in the nodal direction for the YRZ dispersion. These
are given in units of $1/a$ and $t_0a/\hbar$, respectively. 
(b) The real part of the optical conductivity (arbitrary units) 
in the pseudogapped normal
state as a function of $\Omega/t_0$, 
illustrating the Drude and interband components. 
(c) A plot comparing the plasma frequency $\Omega_p^2$ and  the square of the
inverse penetration depth. Also, shown are
the Drude and interband optical spectral weights, $W_D$
and $W_{IB}$ as a function of doping $x$ and the interband spectral weight in
the superconducting state $W_{IB,SC}$. For all of these quantities shown here
we have left out the doping-dependent prefactor $g_t^2$. These curves
are all in units of $e^2t_0/\hbar^2 d$, where $d$ is the $c$-axis
distance
per Cu-O plane.
}
\end{figure}
We begin with a discussion of the validity (or lack thereof) 
 of this rule in the
YRZ model. We will work only at $T=0$ which is simplest and where
the superconducting gap is fully developed and so has its maximum
effect
on charge dynamics. Thus we would expect that, if the presence of
the
pseudogap leads to a violation of Eq.~(\ref{eq:cond7}), it should be
most
noticeable here. In Fig.~\ref{fig2}(c), we plot several of the relevant
quantities as a function of doping, $x$. The solid black dots give 
$\Omega_p^2/8$, {\it i.e.}, the right hand
entry of Eq.~(\ref{eq:cond7}). The change in behavior at $x=0.2$
corresponds to the QCP in our phase
diagram and signals the emergence
of a finite pseudogap. Since we are in the clean limit, and also in
the normal state, $\Omega_p^2/8$ corresponds to $W_D/2$
(green dash-double-dotted curve) in the Fermi liquid region of the phase
diagram $x\ge 0.2$. Below this doping, $\Omega_p^2/8$ has two
contributions
$W_D/2$ from the Drude peak and $W_{IB}$ (long-dashed blue curve)
from the interband transitions. A plot which illustrates
the two contributions
to  ${\rm Re} \sigma(T=0,\Omega)$ is shown in 
Fig.~\ref{fig2}(b). For this plot we used a formula for the
conductivity
which included elastic impurity scattering with quasiparticle
scattering
rate $\eta$ set equal to $0.01t_0$. While the Drude and interband
contributions
are not completely separated, the two distinct contributions remain
clearly identified. The Drude is centered about $\Omega=0$ and the
interband piece is 
shifted to higher energies with an onset just above $\Omega=0.2t_0$. 
The plot is for doping $x=0.14$. This is a case
where the pseudogap is larger than the superconducting gap with 
$\Delta_{\rm pg}/t_0=0.18$ which corresponds roughly to the onset mentioned
above. Returning to Fig.~\ref{fig2}(c), the sum of $W_D/2$
and $W_{IB}$  add up   to $\Omega_p^2/8$
 in the normal state. As we have already mentioned,
in the superconducting state at zero temperature in the clean limit,
the entire Drude condenses into the superfluid and there is no
contribution to the optical spectral weight $\int_{0^+}^\infty {\rm
  Re}\sigma_S(T=0,\Omega)d\Omega$ from this term, only the interband
term remains. Its value as a function of doping is denoted by
$W_{IB,SC}$
and is given by the solid black curve. Comparing this with its normal
state value (long-dashed blue curve), we see that at doping just below
the QCP at $x=0.2$, this contribution is very small and so most of
$W_{IB}$ also
goes into the condensate. However, as $x$ decreases towards the more
underdoped regime, much less of $W_{IB}$ condenses and by $x
\simeq 0.1$, this condensation has stopped. This behavior is expected
and shows that the interband piece is less susceptible to
condensation
into Cooper
pairs than is the Drude
and that this trend increases rapidly as the pseudogap energy
becomes large as compared with twice the superconducting gap energy.
In our phase diagram, this boundary comes at about $\sim 0.1$ doping
as
we would expect.

The dot-double-dashed purple curve gives $1/8\lambda(0)^2$. When this is
added to $W_{IB,SC}$, we obtain, within our
numerical
accuracy, the $\Omega^2_p/8$, {\it i.e.}, the
solid
black circles. Thus we find that to this precision, the YRZ model
also satisfies Eq.~(\ref{eq:cond7}) which is one of our important
results about optical spectral weight distribution
 in addition to our observation that little of the
interband contribution to the conductivity condenses into Cooper
pairs when $\Delta^0_{\rm sc}$ becomes small as compared to the pseudogap
$\Delta^0_{\rm pg}$. Finally, in Fig.~\ref{fig2}(a),
 we show our results for the Fermi momentum in the nodal
direction (solid black curve) as well as the corresponding
Fermi velocity (dashed red curve). Our values agree well with those
given in the work of YRZ and will be needed in a following section. 

\begin{figure}
\includegraphics[width=0.45\textwidth]{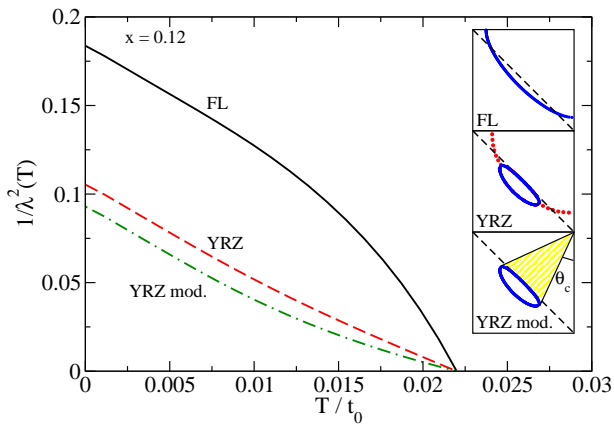}
\caption{\label{fig3} (Color online) The temperature dependence of the
penetration depth shown as $1/\lambda^2(T)$ versus $T/t_0$. Here, the 
$g^2_t$ prefactor, that arises from the definition of the
coherent part of the Green's function, is included. 
Shown are curves for the Fermi liquid
state (FL, solid black curve) and the modification which occurs when a
pseudogap is present in the YRZ model (red dashed curve). The curve
labeled YRZ mod. (green dash-dotted) is a modification of the YRZ
model where the superconducting gap is only non-zero on the Fermi pocket.
The three Fermi surfaces involved are illustrated schematically in the
inset (see text) including the critical angle $\theta_c$.
}
\end{figure}

\section{Results}

We will now present in detail our results and analysis for the penetration
depth. Note that up until now, in our formulas and discussion, we have
not included the Gutzwiller $g_t$ prefactor for weighting only the
coherent part of the Green's function in Eq.~(\ref{eq:G}). Due to the
product of two spectral functions that enters the formulas for the
penetration depth and the optical conductivity 
[Eq.~(\ref{eq:BB}) and Eq.~(\ref{eq:cond1}), respectively], $1/\lambda^2(T)$
and $\sigma(T,\Omega)$ will include a doping-dependent
prefactor of $g_t^2$. We now include it from here
on forward as this factor reflects the highly correlated nature of
the cuprates
and is necessary for a proper comparison with experiment.

In Fig.~\ref{fig3}, we present our results for the temperature
dependence of the penetration depth for $x=0.12$. Three cases are
considered and are to  be compared. The solid black curve labeled
Fermi liquid  is obtained when the pseudogap is set equal to zero
and is for comparison with the short-dashed red curve which
includes the pseudogap and is labeled YRZ. We note two
striking differences between these two sets of results. First,
as a function of temperature, the FL curve shows concave down
behavior. 
This differs slightly from the perhaps more classical curve for a
$d$-wave superconductor  in a model, where the continuum approximation
is used for the band structure and a $\cos(2\theta)$ gap variation is
assumed on the cylindrical Fermi surface, but is mainly due to our use of
a gap ratio of $2\Delta_{\rm sc}^0/k_BT_c=6$ rather than the weak coupling
BCS $d$-wave value of 4.28. The use of a larger value for
 this ratio is in keeping with experimental observation that finds it
of order of 6 (and occasionally reported to be even larger). 
Our results also represent a generalization in which the
superconducting gap varies over the entire Brillouin zone and the
band structure is given in tight-binding with up to third nearest
neighbor hopping. The same gap and bands are used for the red dashed
curve which by contrast shows slight concave upward variation as a
function
of temperature. Here a pseudogap is  included and this has changed
the usual large Fermi surface of Fermi liquid theory which is shown in
one quadrant of the Brillouin zone in the top inset, to the Fermi
surface
shown in the middle inset. Here, the blue curve is the Luttinger
pocket
of the YRZ theory and the red line is its shadow extension which
represents a momentum contour of minimum approach to the
``Fermi surface'' in regions of momentum space where 
a gap exists so there are no true zero energy
states.
This reconstruction of the Fermi surface into a Luttinger pocket,
which by its construction contains exactly $x$ empty states (holes),
has led to a large suppression of the zero temperature penetration
depth as expected in our simplified Eq.~(\ref{eq:pen6}). 
The second
striking feature of these two curves is that they have identical
values of slope, as a function of temperature $T$, out of $T=0$. 
The formation of the Luttinger pockets, as the pseudogap increases in
the underdoped regime of $x$ below the QCP at $x=0.2$ in our phase
diagram, reduces the amount of Fermi surface that is available for 
pairing as in other competing order parameter
scenarios\cite{sharapov},
but leaves the nodal region ungapped. The very low temperature
excitations
out of the ground state are confined to the cone very near zero energy,
which exists in the nodal direction only, but these excitations
do not sample
directly the pseudogap and so the slope retains its Fermi liquid
value.

The final curve in Fig.~\ref{fig3} (green dash-dotted) is a case
where we have cut off the superconducting gap outside the solid angle
$1-4\theta_c/\pi$ which defines the region 
of the
Luttinger pocket, with
$\theta_c$ shown in the lower inset. 
As we expect from the above arguments this does not
change the low temperature slope but does affect the zero temperature
value of the superfluid stiffness which is further reduced over its
Fermi liquid value. Our motivation for applying this cutoff is the
expectation that the superconducting gap will form mainly on the part
of the Fermi surface which remains ungapped. This idea is consistent
with recent ARPES data\cite{arpes} and will be discussed further in
a future paper\cite{jamesarpes}.

\begin{figure}
\includegraphics[width=0.45\textwidth]{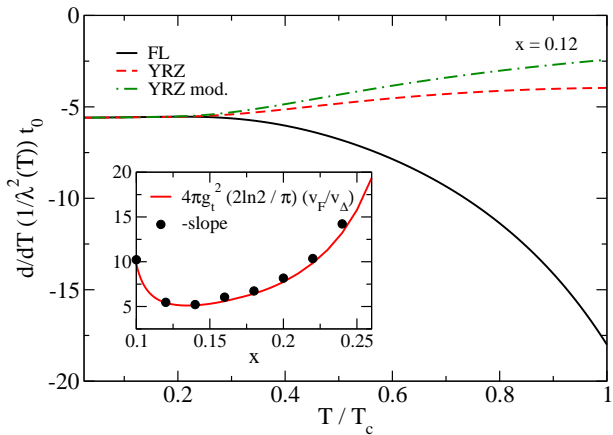}
\caption{\label{fig4} (Color online) Plot of slope of the penetration
depth curves shown in Fig.~\ref{fig3} 
versus $T/T_c$. The inset shows a comparison of
the magnitude of the slope as $T\to 0$ with the formula 
$4\pi g_t^2[2\ln 2/\pi](v_F/v_\Delta)$.
This expression (solid line) is shown 
tested against the numerical data (dots) as a function
of doping $x$.
}
\end{figure}

Figure~\ref{fig4} further emphasizes the insensitivity of 
the low temperature slope of
$1/\lambda^2(T)$ 
 to the formation of the Luttinger pockets
and any cut off in the superconducting gap away from the nodal region.
Shown in Fig.~\ref{fig4} is $t_0d/dT[1/\lambda^2(T)]$ vs $T/T_c$ for the three
cases
illustrated in Fig.~\ref{fig3}. The line labels are the same.
We see perfect agreement between the three curves below $T/T_c\sim
0.3$.
With increasing temperature, the Fermi liquid case shows a downward
trend while by contrast the two other  non-Fermi liquid curves show
the opposite. In the inset to the figure we show our results for the
slope in the YRZ model (solid black dots) as a function of
doping. Also
shown for comparison is the formula 
\begin{equation}
t_0\frac{d}{dT}\biggl[\frac{1}{\lambda^2(T)}\biggr]=4\pi
g_t^2
\frac{2\ln 2}{\pi} \frac{v_F}{v_\Delta},
\label{eq:slopeform}
\end{equation}
where 
$v_\Delta=|\nabla\Delta_{\rm sc}(k)|_{k_F}=(\Delta^0_{\rm
  sc}/\sqrt{2})|\sin(k_{Fx})|$, with all quantities defined in terms of the
units stated in Fig.~\ref{fig2} and elsewhere.
Here, the $|\sin(k_{Fx})|$ factor accounts for the fact that the superconducting
gap amplitude sampled on the Luttinger pocket is somewhat smaller than the
input value $\Delta^0_{\rm sc}$ which enters the phase diagram and corresponds
to the maximum gap in the Brillouin zone. Formula (\ref{eq:slopeform}) 
is derived for
a Fermi liquid with general band structure and we see here that it
also
applies to YRZ theory (Fig.~\ref{fig4}).

\begin{figure}
\includegraphics[width=0.45\textwidth]{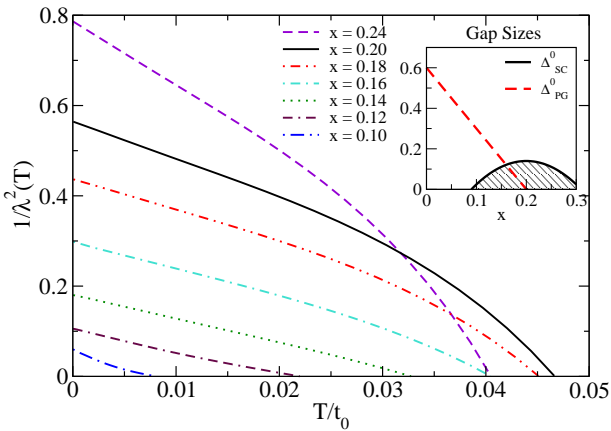}
\caption{\label{fig5} (Color online) 
The temperature dependence of the inverse square
of the penetration depth
for various dopings. The inset shows the values of the
superconducting gap and pseudogap in units of $t_0$ for
varying doping parameter $x$.
}
\end{figure}

In Fig.~\ref{fig5}, we show $1/\lambda^2(T)$ vs $T/t_0$ for different
values  of doping $x$. The phase diagram identifying
the superconducting gap and pseudogap values used is shown for
reference
 in the
inset where we see the QCP at $x=0.2$ corresponding
to the onset of a finite pseudogap which progressively modifies
the large Fermi surface of Fermi liquid theory to smaller Luttinger
pockets. Several overall trends displayed in the $1/\lambda^2(T)$ curves 
shown in this figure are in qualitative agreement with two recent data
sets, one for highly underdoped YBa$_2$Cu$_3$O$_{6+y}$ (YBCO)\cite{broun,sfuubc} 
 and the other for the Bi-based cuprates (Bi:2212)
 Bi$_{2.15}$Sr$_{1.85}$CaCu$_2$O$_{8+\delta}$ and
Bi$_{2.1}$Sr$_{1.9}$Ca$_{0.85}$Y$_{0.15}$Cu$_2$O$_{8+\delta}$ (Ref.~\cite{anukool}).
In addition to the reduction in the $T=0$ value of $1/\lambda^2(T)$,
there is also a trend for the linear temperature dependence out of
$T=0$
to remain to higher values of reduced temperature $T/T_c$. This
contrasts
with the concave downward behavior seen in overdoped and optimally
doped
cases\cite{bonnhardy} as $T_c$ is approached. Such a trend towards linearity
is clearly
seen in the data of Huttema et al.\cite{sfuubc,broun} who note a near linear
dependence of the data almost all the way to $T_c$. However,
they also find, in Fig.~3 of their paper, a turnaround to a $T^2$
law at low temperatures in agreement with the well-known effect of
impurities in $d$-wave superconductors.
While we have not included impurities in our work, we
expect no modifications to this law to arise in the YRZ model and so
agree with these results. However, because the YBCO data is in the
deeply underdoped regime near the bottom of the superconducting dome,
there could be additional effects that become important such as
fluctuations,\cite{lemberger}
 not included here. Hence we turn to the data of Anukool
et al.\cite{anukool} on Bi:2212 and in particular their Fig.~3(b)
which, in agreement with our findings, shows near linear behavior 
over the entire temperature range
considered. Similar behavior
is
also seen in their purer sample (shown in their Fig.~3(a)) although in
that
case the overdoped samples show a low temperature upturn which may
indicate physics not included here. The Bi:2212 data set will be 
examined more closely in what follows as the doping range is more
compatible with the assumptions of the YRZ model.

\begin{figure}
\includegraphics[width=0.45\textwidth]{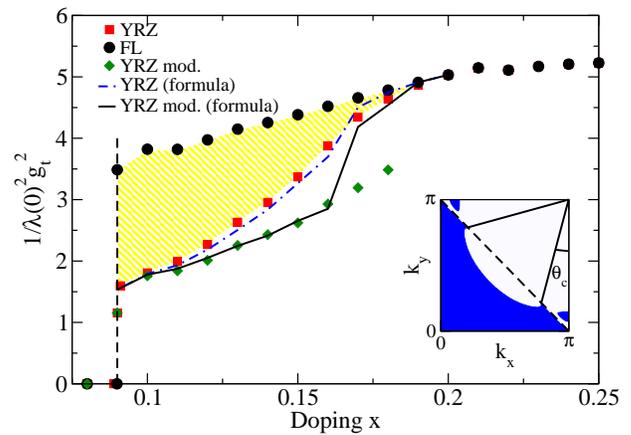}
\caption{\label{fig6} (Color online) The $T=0$ inverse square
penetration depth as a function of doping, comparing the cases
of the Fermi liquid (FL), YRZ and the modified YRZ ansatz.
The shade region emphasizes the reduction in this quantity due to
Fermi surface reconstruction. The curves are for simple analytic
formulas as discussed in the text.
The inset shows the Luttinger
area (shaded blue) for the case of $x=0.18$, illustrating an intermediate
regime of the Fermi surface reconstruction as a function of doping.
This inset also illustrates the critical angle $\theta_c$ for defining
the Fermi pocket, which has become ill-defined in this case.
}
\end{figure}

In Fig.~\ref{fig6}, we show our results for the value of the zero
temperature penetration depth as a function of doping $x$ for three
models. The solid black circles are based on
a Fermi liquid model for the renormalized band which 
 includes the Gutzwiller factors 
of the YRZ theory but no pseudogap. We see a smooth evolution
with increasing superfluid density as doping increases. The solid
red squares are for comparison and include the pseudogap which is finite
below the QCP. A finite $\Delta_{\rm pg}$ leads to Fermi surface reconstruction
as shown in the insets of Fig.~\ref{fig3} where the Luttinger pockets are
seen in the middle and lower frame and are to be compared with the large
Fermi liquid Fermi surface of the top panel. For $x$ just below the 
QCP, the Fermi surface reconstruction can be even more complex than 
in Fig.~\ref{fig3}. An example is shown in the inset of Fig.~\ref{fig6}
for $x=0.18$, where we show Luttinger surface with the dashed curve being
the AFBZ boundary. 
Note the pieces of occupied $\boldsymbol{k}$-space on the other side
of the AFBZ boundary.
In the main frame, the
 solid green diamonds include the pseudogap and, in addition, the superconducting
gap is assumed to be non-zero only on the Luttinger pocket in an attempt to make
the calculations more realistic.
As we saw in Eq.~(\ref{eq:pen6}) with a
simplified continuum model for the band structure, we expect $1/\lambda^2(0)$
to drop as the Fermi surface arc is reduced because the size of the Luttinger
pocket shrinks. The solid black curve corresponds to the
product of the Fermi liquid value of $1/\lambda^2(0)$ with the factor
 $1-4\theta_c/\pi$, the latter 
gives an approximate measure of the ratio of the angle of the 
remaining Fermi arc
compared with that for the Fermi liquid. This curve
 follows the same
general trend as do the results of detailed calculations. 
This shows that Fermi surface reconstruction is responsible for the
drop
in solid green diamonds below the Fermi liquid values (solid black circles).
We note again in this
regard that as the QCP is approached from below ($x<0.2$), there is a region
where the Fermi surface reconstruction is  complex (see insert)
and is not easily characterized by a single arc length. Here, we use the
values previously derived within the context of an application of the YRZ
model to the specific heat\cite{james}. This explains why the solid
black line starts to deviate from the solid green diamonds in the
region near the QCP. We emphasize that the solid green
diamonds include a cut off on the superconducting gap.

As can be seen in the approximate qualitative 
formula (\ref{eq:pen6}), we expect the superconducting gap to affect the zero
temperature condensate, a result which is quite different from the Fermi
liquid case where $1/\lambda^2(0)$ depends only on normal state parameters
and not on the gap. Recently Kopnin and Sonin\cite{kopnin} found a
similar dependence on superconducting gap in the case of graphene
although no superconductivity has yet been reported in this system
 even though carbon
nanotubes have been found to superconduct.\cite{nanotubes}
Returning to Eq.~(\ref{eq:pen6}), the correction factor of
$c\equiv (-4\theta_c/\pi)
(\Delta^{0}_{\rm pg})^2/[(\Delta_{\rm sc}^{0})^2+(\Delta_{\rm
    pg}^{0})^2]$
reduces to $-4\theta_c/\pi$ only when $\Delta_{\rm sc}^0$ is assumed
to be zero when $\Delta^0_{\rm pg}$ is finite, the case described so
far.
If the superconducting gap is allowed to exist over the entire Fermi
surface, the correction factor $c$ will be less than $-4\theta_c/\pi$
and we get the blue dash-dotted curve in Fig.~\ref{fig6} as the
product
of $(1-c)$ times the Fermi liquid value of $1/\lambda(0)^2$ 
[{\it i.e.}, Eq.~(\ref{eq:pen6})]. This
curve agrees well with the solid red squares which are obtained from
the full calculations. It is clear that Fermi surface reconstruction
leads to a strong reduction of the zero temperature superfluid
density as a result of the loss of ungapped states on the Fermi surface.

\begin{figure}
\includegraphics[width=0.45\textwidth]{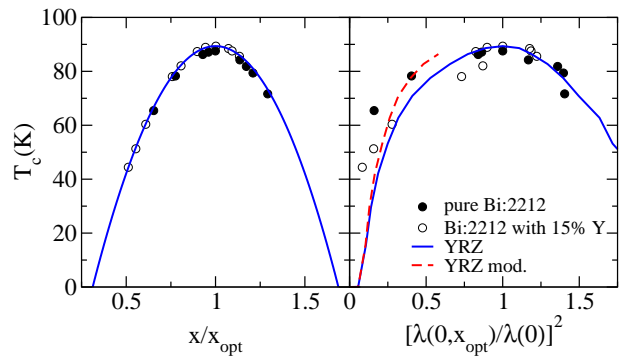}
\caption{\label{fig7} (Color online) 
$T_c$ vs doping
and vs the superfluid density normalized to the optimal value in the YRZ model.
The solid blue curve is the result of the YRZ model 
and the red dashed curve is the modified YRZ model as discussed in the
text.
Comparison has been made 
with the data of Anukool et al.\cite{anukool} with the Bi:2212 (solid dots) and Bi(Y):2212
(open dots).
}
\end{figure}

We now turn to experiment to assess the validity of the theoretical
model due to YRZ. Recent results as a function of doping
on the penetration depth of Bi-based
cuprates (Bi:2212) have been published by Anukool et al.\cite{anukool}.
They give results for a range of doping from underdoped to overdoped and for
$T_c$'s reduced by up to 50\%. Because their optimal doping is at $p=0.16$
and our model uses $x=0.2$, we shifted the data set doping values by 0.04
in order to match their dome with ours. Normalizing to the optimal doping,
we then find that their $T_c$ curve matches our $T_c$ dome if we
convert our energy gap dome to one for $T_c$ by using a value of the gap ratio
$2\Delta^0_{\rm sc}/k_BT_c=6$ and then taking $t_0=165$ meV (our energy units
have all been in units of $t_0$). With this we find a good match to the
experimental $T_c$ versus doping curve shown in Fig.~\ref{fig7}. This now
fixes our parameters for the rest of our work. In Fig.~\ref{fig7}
(right hand
frame) we show the classic Uemura plot of $T_c$ versus $1/\lambda^2(0)$,
where we have normalized our results and the data to the value at optimal
doping. The agreement with the data is excellent and on the underdoped
side, the data appears to favor more the modified YRZ calculation.
\begin{figure}
\includegraphics[width=0.45\textwidth]{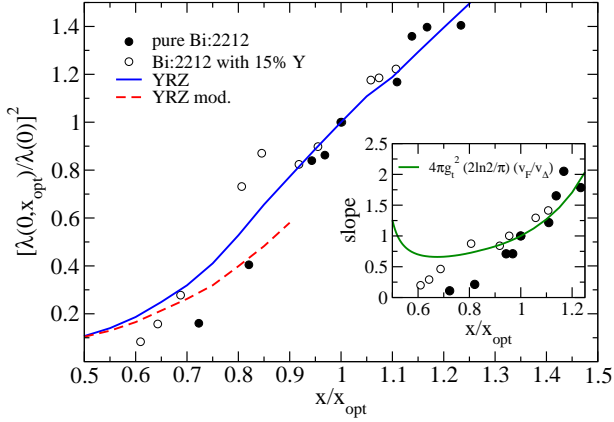}
\caption{\label{fig8} (Color online) Comparison of the
superfluid density at $T=0$ with the Bi:2212 (solid dots) and Bi(Y):2212
(open dots)
data of Anukool et al.\cite{anukool}. This is
shown as a function of doping scaled to
the optimal doping value. 
The solid blue curve is the result of the YRZ model 
and the red dashed curve is the modified YRZ model as discussed in the
text. The inset shows the slopes of the data as $T\to 0$ in comparison
with the analytic formula discussed in Fig.~\ref{fig4}.
}
\end{figure}
Furthermore, in Fig.~\ref{fig8}, we show the data compared with the inverse
penetration
depth squared versus doping and once again the data favors the modified 
YRZ curve. What is clear here is that the data agrees well with the Fermi liquid
curve above optimal doping and on the underdoped side it follows the YRZ
curve which includes the reconstructed Fermi surface. Indeed, the drop
in the data below optimal doping can be taken as possible indirect experimental
evidence for Fermi surface reconstruction as the Fermi liquid curve would have
been higher (as discussed in relation to Fig.~\ref{fig6}). All is not ideal,
however.
We have used a normalized quantity here. 
In absolute value, we find that the superfluid density is off by about a
factor of 3-4. Nonetheless, this can be rectified as the penetration depth
depends only on $t_0$  and so one could increase the value of $t_0$
to account for this. This would then require an adjustment of the gap ratio
to keep the good agreement with the $T_c$ dome of Fig.~\ref{fig7},
but we find that it would be unrealistically large. 
Another possibility is to
note that our estimate of the Fermi velocity in the YRZ model at optimum
is low by a factor of two. The superfluid density shown here goes as
the square of the velocity and this could possibly correct the situation.
A related issue is that 
the band structure used in the YRZ paper is for a very different compound
than that of Bi:2212 and so this could change some of these quantitative
numbers. We
wish to stress that the merit of this theory should be seen in its
ability to give qualitative insight into the pseudogap phase and the
good agreement that we find here is very encouraging given the lack of
detailed parameter fitting to this particular material. Also shown in
Fig.~\ref{fig8} is an inset plotting the experimental slope of the
inverse square
penetration depth curves for $T\to 0$. Again this has been normalized to
the optimal doping value and we find good agreement with the analytic
formula discussed for the slope in Fig.~\ref{fig4}. An 
analysis based on ARPES
has given somewhat different results.\cite{norman} It should be noted
that we extracted the slope values ourselves and so this is not a rigorous
representation of the data and indeed the temperature-dependence of some
of the experimental curves was unusual in a few cases, but we took the lowest
temperature value of the slope in any event. The optimal doping curve
was in this category as it showed an upturn at low temperature in the
case of the pure sample and so its slope value and a few of the others
could be revised in the hands of experimental experts. However,
the inset of Fig.~\ref{fig8} (as well as the other comparisons
of Figs.~\ref{fig7} and \ref{fig8}) serves to illustrate that the Gutzwiller
prefactor for the coherent part of the Green's function, leading to
a prefactor of $g_t^2$,  appears to be essential
in giving the correct trend of the data with doping. This points to
very strong correlations in these systems.

\begin{figure}
\includegraphics[width=0.45\textwidth]{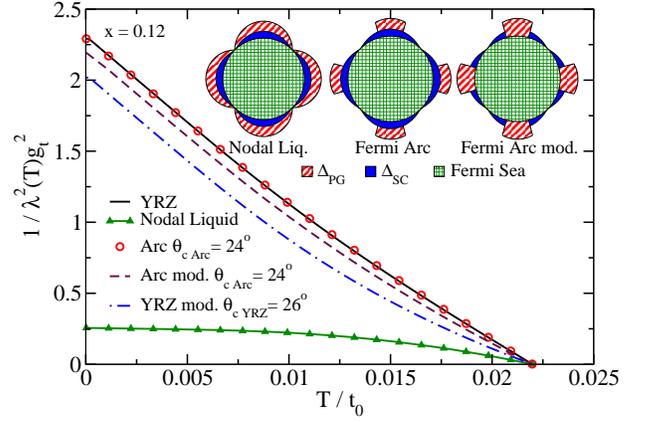}
\caption{\label{fig9} (Color online) 
A comparison of the effect that various models have on the square of
the inverse
penetration depth as a function of temperature.
The schematic diagrams illustrate the momentum dependence
of the two energy gaps for the Fermi arc model
and its modification, relative to the nodal liquid case.
}
\end{figure}
In Fig~\ref{fig9},
we give results which show the relationship of the YRZ model
with two other prominent models. Both involve the assumption that the
pseudogap acts on the Fermi surface. This corresponds to taking the
limit of $\xi^0_{\boldsymbol{k}}\to \xi_{\boldsymbol{k}}$ in our
Eq.~(\ref{eq:yrzpen}) for the penetration depth, {\it i.e.}, to 
replacing the AFBZ energy by the carrier dispersion curve.
As we saw,
this greatly simplifies the equation and reduces it to the familiar
form of Eq.~(\ref{eq:bpen}) for a BCS superconductor
with one modification. While in
Eq.~(\ref{eq:bpen}) the superfluid density
 remains directly proportional to a factor
of $\Delta^2_{\rm sc}$ and so manifestly vanishes in the normal state,
the energy $E_S=\sqrt{\xi_{\boldsymbol{k}}^2+\Delta^2_{\rm sc}+\Delta^2_{\rm pg}}$
involves the sum of the squares of $\Delta_{\rm sc}$ and $\Delta_{\rm pg}$
rather than the square of $\Delta_{\rm sc}$ alone.
It is only through this factor that the pseudogap enters in this
simplification. The Fermi arc model\cite{james} corresponds to taking
 $\Delta_{\rm pg}$ non-zero only on an arc centered about the antinodal
direction, leaving an ungapped region centered around the nodal direction
(on the Luttinger pocket) as illustrated by the center
insert in Fig.~\ref{fig9}. One can construct realistic
models\cite{storey}
 for the momentum variation
of the pseudogap on the Fermi surface. For simplicity, here we retain its
cosine form $\Delta_{\rm pg}=\frac{\Delta_{\rm pg}^{0}(x)}{2}(\cos
k_xa -\cos k_ya)$ but cut it off on an arc defined by the angle $\theta_c$
previously introduced. Taking $\theta_{c,\rm arc}=24^\circ$, chosen to get
a best fit to the penetration depth curve of the YRZ model (solid black curve),
one obtains the red open circles which overlap the black solid curve almost
perfectly. Thus with the right choice of $\theta_{c,\rm arc}$,
one can get an almost perfect match between the two models. If we further
assume that the superconducting gap is present only on the ungapped
(by the pseudogap that is, see right hand inset) 
part of the Fermi surface, we obtain the 
dashed curve denoted Arc mod. This has pushed the penetration depth down
by about the same amount as we saw in Fig.~\ref{fig3} for the full YRZ
calculation when we included a superconducting gap modification in this
model. 
The choice of $\theta_{c,arc}$ in Fig.~\ref{fig9}
is important as can be seen from the dash-dotted
blue curve which is also based on an arc model but with a different cut off model
$\theta_{c,\rm YRZ}=26^\circ$, a value obtained from the size of the arc
subtended by the actual Luttinger pocket of the YRZ theory. While this choice
decreases further the magnitude of the superfluid density because there is
even less ungapped arc, it does not change the qualitative behavior obtained.

Finally, the solid green curve with triangles is our result for the nodal
liquid which corresponds to lifting the cut off on the pseudogap.
In this case an effective gap of $\sqrt{\Delta_{\rm sc}^2+\Delta_{\rm pg}^2}$
replaces the usual superconducting gap in the standard Eq.~(\ref{eq:bpen})
for $1/\lambda^2(T)$ with one critical difference. In the numerator
of Eq.~(\ref{eq:bpen}), it is still $\Delta_{\rm sc}^2$ which remains and not
$\Delta_{\rm sc}^2+\Delta_{\rm pg}^2$. Since we have assumed the same
momentum dependence for the superconducting gap and pseudogap, we can replace
the explicit $\Delta_{\rm sc}^2$ factor in Eq.~(\ref{eq:bpen}) by
$\Delta_{\rm sc}^2+\Delta_{\rm pg}^2$ and take out the factor of 
$(\Delta_{\rm sc}^0)^2/[(\Delta_{\rm sc}^0)^2+(\Delta_{\rm pg}^0)^2]$
to compensate for this. Thus the equation for the penetration depth becomes
the standard one for an effective superconducting gap of 
$\sqrt{\Delta_{\rm sc}^2+\Delta_{\rm pg}^2}$ but with the difference that a
factor of the square of superconducting-to-effective-gap amplitude 
is to multiply the entire expression as we see explicitly in our simplified
expression (\ref{eq:pen4}). While these simplifications allow us to obtain simple
analytic expressions, we see that this limit fails to give quantitative
results when compared with YRZ. Compare the green curve with triangles with
the solid black curve in Fig~\ref{fig9}. In fact one should not expect
the two models to agree. The nodal liquid is the limit when the size of the
Luttinger pocket tends to zero. But this is never reached inside the
superconducting dome. In our calculations at a doping where $T_c$
has been depressed to zero, there remains a sizable part of the Fermi
surface which is ungapped. It is not surprising then
that the nodal liquid should show qualitative behavior not part of the arc
model, such as, 
a superfluid density at $T=0$ which decreases by a factor of the 
square of the ratio of superconducting to pseudogap.
The slope of the linear-in-$T$ variation
at low temperature also becomes inversely proportional to 
$\sqrt{\Delta_{\rm pg}^2+\Delta_{\rm sc}^2}$
rather than to $\Delta_{\rm sc}$ and so becomes very flat.

\section{Conclusions}

YRZ have provided a simple model for the coherent part of the
charge carrier Green's function which applies below a quantum critical
point characterizing the beginning of
pseudogap formation. As the Mott-Hubbard transition
to an insulating state is approached with decreasing doping, the magnitude
of the pseudogap increases, the bands narrow and the weight of the
coherent piece of the Green's function decreases according to
well-defined Gutzwiller factors which account for correlation effects.
With increasing pseudogap magnitude, the Fermi surface reconstructs. It
goes from the large Fermi surface of Fermi liquid theory to ever smaller
Luttinger pockets. This leads directly to a reduction of the value of the
zero temperature inverse squared penetration depth because there are fewer
ungapped states which are available to form the condensate. We show that
this reduction is roughly proportional to the ratio of the
remaining arc length, defined
by the Luttinger pocket, to the full length of the corresponding Fermi liquid
surface. On the other hand, and in sharp contrast, the coefficient of the
linear-in-temperature term of $1/\lambda^2(T)$ at small $T$ remains
largely unaffected because this quantity depends only on the available
very low energy excitations and these are confined to the vicinity of the
nodal Dirac points. This region is not importantly affected by pseudogap
formation and implied Fermi surface reconstruction.

A comparison of our results with recent experimental data on Bi:2212
gives good qualitative agreement and demonstrates the importance of the
strong dependence on doping of the coherence weight in the YRZ model which
derives from strong correlation effects. While this is a simple
multiplicative factor, it accounts for a significant part of the 
reduction in superfluid density as the end of the superconducting dome is
approached in the deeply underdoped region of the phase diagram.
An additional reduction is due to Luttinger pocket formation which
starts
at the doping associated with the quantum critical point and
this effect provides a more abrupt change in magnitude which should be
measurable as a signature of Fermi surface reconstruction.

The well-known Ferrell-Glover-Tinkham sum rule of conventional
superconductivity was found to apply here as well. The optical weight lost on entering
the superconducting state below $T_c$ reappears in its entirety in the
superfluid fraction.

Comparison of our results with those based on an arc model, with 
pseudogap placed on the Fermi surface rather than on the AFBZ of the
YRZ model, can be made to agree very well if the arc length on which
the pseudogap is assumed to be non-zero is appropriately chosen to obtain
a best fit with the YRZ case. The nodal liquid concept corresponds to
the limit when the Fermi surface is fully gapped by the pseudogap except
for the nodal points. This limit is never reached in YRZ theory
because the size of the Luttinger pocket remains quite significant at the
doping which corresponds to the end of the superconducting dome. 
Nevertheless, the nodal liquid limit remains valuable because it yields
analytic results which can provide useful insight into the deeply 
underdoped case.

\begin{acknowledgments}
This work has been supported by NSERC of Canada 
and by the Canadian Institute for Advanced Research (CIFAR).
\end{acknowledgments}


\begin{thebibliography}{99}

\bibitem{timusk} T. Timusk and B. Statt,
Rep. Prog. Phys. {\bf 62}, 61 (1999).

\bibitem{carbottenature} See J.P. Carbotte, E. Schachinger, and D. N. Basov,
Nature (London) {\bf 401}, 354 (1999) and references therein.

\bibitem{inversion} E. Schachinger and J. P. Carbotte,
Phys. Rev. B {\bf 62}, 9054 (2000).

\bibitem{laughlin} S. Chakravarty,
R. B. Laughlin, D. K. Morr, and C. Nayak,
Phys. Rev. B {\bf 63}, 094503 (2001).

\bibitem{emery} V. J. Emery and S. A. Kivelson,
Nature (London) {\bf 374}, 434 (1995).

\bibitem{randeria} M. Randeria, N. Trivedi, A. Moreo, and R. T. Scalettar,
Phys. Rev. Lett. {\bf 69}, 2001 (1992).

\bibitem{alvarez} G. Alvarez and E. Dagotto,
Phys. Rev. Lett. {\bf 101}, 177001 (208).

\bibitem{levin} Q. Chen, K. Levin, and I. Kosztin,
Phys. Rev. B {\bf 63}, 184519 (2001).

\bibitem{hufner} S. H\"ufner, M. A. Hossain, A. Damascelli,
and G. A. Sawatzky,
Rep. Prog. Phys. {\bf 71}, 062501 (2008).

\bibitem{HF} H. v. L\"ohneysen, A. Rosch, M. Vojta, and P. W\"olfe.
Rev. Mod. Phys. {\bf 79}, 1015 (2007).

\bibitem{louis} L. Taillefer, arXiv:0901.2313 and references therein.

\bibitem{arpesarcs} For example, 
A. Kanigel, U. Chatterjee, M. Randeria, M. R. Norman,
S. Souma, M. Shi, Z. Z. Li, H. Raffy, and J. C. Campuzano,
Phys. Rev. Lett. {\bf 99}, 157001 (2007); 
A. Kanigel, M. R. Norman, M. Randeria, U. Chatterjee, S. Souma,
A. Kaminski, H. M. Fetwell, S. Rosenkranz, M. Shi,T. Sato, T. Takahashi,
Z. Li, H. Raffy, K. Kadowaki, D. Hinks, L. Ozyuzer, and J. C. Campuzano,
Nat. Phys. {\bf 2}, 447 (2006); T. Kondo, R. Khasanov, T. Takluchi, 
J. Schmalian, and A. Kaminski,
Nature (London) {\bf 457}, 296 (2009).

\bibitem{zhou} J. Meng, G. Liu, W. Zhang, L. Zhao, H. Liu, X. Jia,
D. Mu, S. Liu, X. Dong, W. Lu, G. Wang, Y. Zhou, Y. Zhu, X. Wang,
Z. Xu, C. Chen, and X. J. Zhou,
arXiv:0906.2682

\bibitem{hwang} J. Hwang, J. P. Carbotte, and T. Timusk,
EPL {\bf 82}, 27002 (2008).

\bibitem{storey} J. G. Storey, J. L. Tallon, G. V. M. Williams,
and J. W. Loram, 
Phys. Rev. B {\bf 76}, 060502(R) (2007). 

\bibitem{davis} Y. Kohsaka, C. Taylor, P. Wahl, A. Schmidt,
Jhinhwan Lee, K. Fujita, J. W. Alldredge, Jinho Lee,
K. McElroy, H. Esaki, S. Uchida, D.-H. Lee, and J. C. Davis, 
Nature (London) {\bf 454}, 1072 (2008).

\bibitem{yrz} K.-Y. Yang, T. M. Rice and F. C. Zhang
Phys. Rev. B {\bf 73}, 174501 (2006).

\bibitem{anderson} P. W. Anderson,
Science {\bf 235}, 1196 (1987).

\bibitem{belen} B. Valenzuela and E. Bascones,
Phys. Rev. Lett. {\bf 98}, 227002 (2007).

\bibitem{emilia} E. Illes, E. J. Nicol, and J. P. Carbotte,
Phys. Rev. B {\bf 79}, 100505(R) (2009).

\bibitem{yrzarpes} K.-Y. Yang, H. B. Yang, P. D. Johnson, T. M. Rice and F. C. Zhang,
EPL {\bf 86}, 37002 (2009).

\bibitem{james} J. P. F. LeBlanc, E. J. Nicol, and J. P. Carbotte,
Phys. Rev. B {\bf 80}, 060505(R) (2009).

\bibitem{bonnhardy} W. N. Hardy, D. A. Bonn, D. C. Morgan, R. Liang, and 
K. Zhang,
Phys. Rev. Lett. {\bf 70}, 3999 (1993).

\bibitem{uemura} Y. Uemura et al.,
Phys. Rev. Lett. {\bf 62}, 2317 (1989).

\bibitem{leeandwen} P. A. Lee and X.-G. Wen,
Phys. Rev. Lett. {\bf 78}, 4111 (1997).

\bibitem{franz} D. E. Sheehy, T. P. Davis, and M. Franz,
Phys. Rev. B {\bf 70}, 054510 (2004).

\bibitem{branch1995} D. Branch and J. P. Carbotte,
Phys. Rev. B {\bf 52}, 603 (1995).

\bibitem{tomlinson} P. G. Tomlinson and J. P. Carbotte,
Phys. Rev. B {\bf 13}, 4738 (1976).

\bibitem{leung1} H. K. Leung, J. P. Carbotte, D. W. Taylor, and C. R. Leavens,
Can. J. Phys. {\bf 54}, 1585 (1976).

\bibitem{leung2} H. K. Leung, J. P. Carbotte, and C. R. Leavens,
J. Low Temp. Phys. {\bf 24}, 25 (1976).

\bibitem{branch} D. Branch, Ph.D. thesis, McMaster University, 1997 (unpublished)

\bibitem{radtke} R. J. Radtke, V. N. Kostur, and K. Levin,
Phys. Rev. B {\bf 53}, R522 (1996).

\bibitem{ewald} E. Schachinger, J. P. Carbotte, and F. Marsiglio,
Phys. Rev. B {\bf 56}, 2738 (1997).

\bibitem{donovan1} C. O'Donovan and J. P. Carbotte,
Phys. Rev. B {\bf 52}, 16208 (1995).

\bibitem{donovan2} C. O'Donovan and J. P. Carbotte,
Phys. Rev. B {\bf 52}, 4568 (1995).

\bibitem{donovan3} C. O'Donovan and J. P. Carbotte,
Physica C {\bf 252}, 9433 (1996).


\bibitem{misawa} S. Misawa,
Phys. Rev. B {\bf 51}, 11791 (1995).

\bibitem{FGT1} R. A. Ferrell and R. E. Glover,
Phys. Rev. {\bf 109}, 1398 (1958).

\bibitem{FGT2} M. Tinkham and R. A. Ferrell,
Phys. Rev. Lett. {\bf 2}, 331 (1959).

\bibitem{sharapov} S. G. Sharapov and J. P. Carbotte,
Phys. Rev. B {\bf 73}, 094519 (2006).

\bibitem{arpes} T. Kondo, R. Khasanov, T. Takeuchi, J. Schmalian,
and A. Kaminski,
Nature (London) {\bf 457}, 296 (2009).

\bibitem{jamesarpes} J. P. F. LeBlanc, J. P. Carbotte, and
  E. J. Nicol, (unpublished).


\bibitem{broun} D. M. Broun, W. A. Huttema, P. J. Turner, S. \"Ozcan,
B. Morgan, R. Liang, W. N. Hardy, and D. A. Bonn, 
Phys. Rev. Lett. {\bf 99}, 237003 (2007).

\bibitem{sfuubc}  W. A. Huttema, J. S. Bobowski, P. J. Turner, 
R. Liang, W. N. Hardy, D. A. Bonn, and D. M. Broun, 
Phys. Rev. B {\bf 80}, xxx (2009).

\bibitem{anukool} W. Anukool, S. Barakat, C. Panagopoulos, and 
J. R. Cooper, 
Phys. Rev. B {\bf 80}, 024516 (2009).

\bibitem{lemberger} I. Hetel, T. R. Lemberger, and M. Randeria,
Nature Phys. {\bf 3}, 700 (2007).

\bibitem{kopnin} N. B. Kopnin and E. B. Sonin, 
Phys. Rev. Lett. {\bf 100}, 246808 (2008).

\bibitem{nanotubes} Z. K. Tang, Lingyun Zhang, N. Wang, X. X. Zhang,
G. H. Wen, G. D. Li, J. N. Wang, C. T. Chan, and Ping Sheng,
Science {\bf 292}, 2462 (2001).

\bibitem{norman} J. Mesot, M. R. Norman, H. Ding, M. Randeria,
J. C. Campuzano, A. Paramekanti, H. M. Fretwell, A. Kaminski,
T. Takeuchi, T. Yokoya, T. Sato, T. Takahashi, T. Mochiku,
and K. Kadowaki,
Phys. Rev. Lett. {\bf 83}, 840 (1999).

\end{thebibliography}
\end{document}